\documentclass[reqno,12pt]{article}
\usepackage{amsmath,amsfonts,amssymb,amsthm,amstext,amscd,eucal}
\usepackage[all]{xy}
\newcommand{\cO}{{\cal O}}

\newcommand{\be}{\begin{equation}}
\newcommand{\ee}{\end{equation}}

\newcommand{\bse}{\begin{subequations}}
\newcommand{\ese}{\end{subequations}}
\newcommand{\bea}{\begin{eqnarray}}
\newcommand{\eea}{\end{eqnarray}}
\newcommand{\ba}{\begin{array}}
\newcommand{\ea}{\end{array}}

\makeatletter \@addtoreset{equation}{section}

\makeatother
\newcommand{\ord}[1]{\mbox{{$\cO$}}\left({#1}\right)}

\def\<{\langle}
\def\>{\rangle}

\def\SW{Seiberg-Witten\ }

\def \th {\theta^{\mu\nu}}

\def\nc{noncommutative\ }

\def\SWmap{Seiberg-Witten map }

\def\sm{Standard Model }
\def\Uo{$U_{\star}(1)$ }

\def\sp{$\star$-product}

\def\nbyn {$N\!\times\! N$}

\begin{document}
\baselineskip 18pt%

\begin{titlepage}
\vspace*{1mm}%
\hfill%
\vbox{
    \halign{#\hfil \cr
           \;\;\;\;\;\;\;\;\;\;\;\;\;\;IPM/P-2009/024 \cr
           \cr
           } 
      }  
\begin{center}
{{\Large {\bf  \textsl{Can Seiberg-Witten Map Bypass\\
\vskip3mm
 Noncommutative Gauge Theory No-Go Theorem?}
}}}%
\vspace*{10mm}


{\bf \large{ M. Chaichian$^{\dagger}$, P. Pre\v{s}najder$^{\S}$, M.
M. Sheikh-Jabbari$^{\ddag}$, {{A. Tureanu}}$^{\dagger}$}}

\end{center}
\begin{center}
\vspace*{0.4cm} {\it { $^{\dagger}$ Department of Physics,
University of Helsinki\\
and Helsinki Institute of Physics,
P.O. Box 64, FIN-00014 Helsinki, Finland\\
$^\S$Department of Theoretical Physics, Comenius University,\\
Mlynsk\'{a} dolina, SK-84248 Bratislava,
Slovakia \\
$^{\ddag}$ School of Physics, Institute for Research in Fundamental Sciences (IPM)\\
P.O.Box 19395-5531, Tehran, IRAN}}\\
\centerline{\tt masud.chaichian@helsinki.fi,
presnajder@fmph.uniba.sk, jabbari@theory.ipm.ac.ir,} \centerline{\tt
anca.tureanu@helsinki.fi}

\vspace*{.5cm}
\end{center}

\begin{center}{\bf Abstract}\end{center}
\begin{quote}
There are strong restrictions on the possible representations and in
general on the matter content of gauge theories formulated on
noncommutative Moyal spaces, termed as noncommutative gauge theory
no-go theorem. According to the no-go theorem \cite{no-go}, matter
fields in the noncommutative $U(1)$ gauge theory can only have $\pm
1$ or zero charges and for a generic noncommutative $\prod_{i=1}^n
U(N_i)$ gauge theory matter fields can be charged under at most two
of the $U(N_i)$ gauge group factors. On the other hand, it has been
argued in the literature that, since a noncommutative $U(N)$ gauge
theory can be mapped to an ordinary $U(N)$ gauge theory via the
Seiberg-Witten map, seemingly it can bypass the no-go theorem. In
this note we show that the Seiberg-Witten map  \cite{SW} can only be
consistently defined and used for the gauge theories which respect
the no-go theorem. We discuss the implications of these arguments
for the particle physics model building on noncommutative space.

\end{quote}

\end{titlepage}
\textwidth 16.5cm%

\section{Introduction}

Motivated by string theory considerations  (e.g. see \cite{SW}), the
possibility of having a noncommutative (NC) Moyal spacetime, with
the coordinate operators satisfying the commutation relations:
\[
[\hat x^\mu,\hat x^\nu]=i\th,
\]
with $\th$ being a given constant tensor, as physical spacetime, has
been under intense study. Using the Weyl-Moyal correspondence, there
is a simple recipe for constructing quantum field theories on NC
Moyal spaces: replace the usual product between the fields in the
action by the Moyal star product,
\begin{eqnarray}\label{star}%
\big(f \star g\big)(x)&
=&e^{\frac{i}{2}\theta^{\mu\nu}\frac{\partial}{\partial
    x^\mu}\frac{\partial}{\partial y^\nu}}f(x)g(y)|_{y= x}\\
& = &f(x)g(x)+\frac{i}{2}\theta^{\mu\nu}\partial_\mu
f(x)\partial_\nu
g(x)-\frac{1}{8}\theta^{\mu\nu}\theta^{\sigma\tau}\partial_\mu\partial_\sigma
f(x)\partial_\nu\partial_\tau g(x)+\ord{\theta^3},\nonumber
\end{eqnarray}
where $\theta^{\mu\nu}=-\theta^{\nu\mu}$.

For obvious particle physics reasons, as well as string theoretical
motivations, gauge field theories on NC Moyal space (NC gauge
theories) have also been studied. Due to the fact that the Moyal
star product \eqref{star} is noncommutative, there are severe
restrictions on the possible gauge groups and matter contents in NC
gauge theories. For example, using the above-mentioned recipe one
can easily see that the only possible gauge group is the
noncommutative $U(N)$, denoted by $U_\star(N)$, where the gauge
fields are $N\times N$-valued Hermitian matrices. As will be briefly
reviewed in the next section, for similar reasons the matter fields
in NC gauge theories can only be in three representations of $U(N)$:
fundamental, antifundamental or adjoint. For the particular case of
the $U_\star(1)$ gauge theory, the NC QED, due to the noncommutative
nature of the products of fields, the allowed charges are limited to
$0, \pm 1$ \cite{Hayakawa}. Moreover, when we have a general gauge
group which is composed of a product of several $U_\star(N_i)$
factors, the matter fields can carry charges under at most two of
the $U_\star(N_i)$ factors. These restrictions on NC gauge theories
were crystalized in the NC gauge theory no-go theorem \cite{no-go}
(see also \cite{Terashima,Carmelo}) and used constructively in
building a noncommutative version of the Standard Model \cite{NCSM}.

On the other hand, motivated by string theory analysis that a NC
Moyal space arises from the worldvolume of a D-brane in a constant
background Neveu-Schwartz two-form field $B$, in \cite{SW} it was
proposed that a NC gauge field theory (which can be thought of as
the low energy effective theory of open strings attached to the
D-brane) should also have a description in terms of a gauge theory
on a commutative spacetime. This ordinary gauge theory will,
however, have a complicated action. Based on the intuition coming
from string theory, Seiberg and Witten proposed a map, the
Seiberg-Witten map \cite{SW}, between ordinary and noncommutative
gauge theories and their fields and gauge transformations. This map
can in principle be constructed explicitly, using the defining
equation, as a systematic expansion in powers of $\th$. A short
review on the \SW map will be presented in Appendix
\ref{SW-review-appendix}.

It has been argued that  the Seiberg-Witten map, which relates a \nc
gauge theory to an ordinary one, paves the way for constructing the
\nc version of gauge theories based on generic Lie algebras with
matter fields in generic representations
\cite{Madore,Wess-enveloping,Wess-general-NCgauge-theory}, thus
circumventing the restrictions discussed in \cite{no-go}. In
particular, it has been argued that one can have NC($su(N)$) gauge
theories, as well as having NC $U(1)$ theory with arbitrary charges
\cite{Wess-enveloping,Wess-general-NCgauge-theory,Wess-NCSM}. The
NC($su(N)$) case is based on a construction relying on the notion of
enveloping algebra, which will be reviewed in Appendix
\ref{enveloping-appendix}. These ideas have then been employed for
the construction of a noncommutative Standard Model
\cite{Wess-NCSM}. Like the \SWmap itself, this model has been
defined as a series expansion in powers of $\th$.

In this letter we study the question:\\
\centerline{ \emph{Can the \SWmap bypass the noncommutative gauge
theory no-go theorem?}}\\ We shall argue that indeed the use of the
\SWmap is limited only to the cases which respect the NC gauge
theory no-go theorem. Although the no-go theorem \cite{no-go}, as it
stands, does not apply to the NC($su(N)$) constructed via the use of
the enveloping-algebra-valued gauge fields and gauge
transformations, we show that the very definition of the \SW map
still forbids matter fields being in fundamental representation of
more than one NC($su(N)$) factor (or, equivalently, forbids matter
fields from being charged under more than two noncommutative gauge
groups/algebras).

This letter is organized as follows. In Section
\ref{no-go-vs.SW-section}  we show that the \SW map can only be used
for the cases which respect the \nc gauge theory no-go theorem. We
also discuss that our results are supported by the string theory
intuition and expectations. In Section \ref{discussion-section} we
discuss the implications of our results for the \nc particle physics
model building. To be self-contained, in two Appendices we review
the \SW map, its definition and consistency conditions, as well as
the construction of NC($su(N)$) gauge theories based on the notion
of enveloping algebra and the \SW map.

\section{Noncommutative gauge theory no-go theorem and the
Seiberg-Witten map}\label{no-go-vs.SW-section}

The no-go theorem for noncommutative gauge theories, as formulated
in \cite{no-go}, states that: 1) the local NC $u(N)$ \emph{algebra},
denoted as $u_\star(N)$, only admits the irreducible $N\times N$
matrix representation. Hence the gauge fields are in $N\times N$
matrix form, while the matter fields \emph{can only be} in
fundamental/antifundamental, adjoint or singlet states; 2) for any
gauge group consisting of several simple-group factors, the matter
fields can transform nontrivially under \emph{at most two} NC gauge
group factors, in bi-fundamental representaion. In other words, the
matter fields can not carry more than two NC gauge group charges.

The \SW map, on the other hand, relates the NC gauge field theories
to ordinary gauge field theories, which do not have the restrictions
on the representations of matter fields implied by the no-go
theorem. In this section we re-examine, in view of the \SW map, the
two parts of the no-go theorem, in particular the charge
quantization problem and the property of not-more-than-two charges.

\subsection{Ordinary and noncommutative gauge
theories}\label{ord_vs_NC}

Let us recall some basic facts about ordinary and noncommutative
gauge theories and introduce the notations. In what follows, the
hatted functions are multiplied with the $\star$-product
\eqref{star}, while the unhatted quantities -- with ordinary product
of functions. The matrix part of the $u_\star(N)$ algebra is
generated by the $N\times N$ Hermitian matrices $T^a,\ a=1,2,\ldots,
N^2-1$, with the Lie-algebra structure $[T^a,T^b]=if^{abc}T^c$,
normalized  as $Tr(T^aT^b)= {1\over 2} \delta^{ab}$, to which we add
$T^0={1\over\sqrt{2N}}{\bf 1}_{N\times N}$.

In the ordinary $U(N)$ Yang-Mills theory, the gauge fields%
\be\label{un-g.f.ord}%
A_{\mu}(x)=\sum_{a=0}^{N^2-1}\ A_{\mu}^a(x) T^a %
\ee
transform under the infinitesimal gauge transformation
$\delta_\Lambda$ as:
\be\label{g.t.ord}%
 A_{\mu}\to\  A'_{\mu}= A_{\mu}+\delta_\Lambda
A_\mu= A_{\mu} +
\partial_{\mu} \Lambda + ig [\Lambda, A_\mu]\ ,\ \ \
\; \Lambda=\sum_{a=0}^{N^2-1}\ \Lambda^a(x)\,T^a\,\in u(N)\,, %
\ee
and the corresponding field strength
\be\label{f-s.ord} F_{\mu\nu}=\partial_{[\mu}A_{\nu]} + i g[
A_{\mu},A_{\nu}]\ , \ee
transforms covariantly
\be F_{\mu\nu}\to\ F'_{\mu\nu}= F_{\mu\nu} +i g [\Lambda,
F_{\mu\nu}]\ . \ee
In the above, $g$ is the gauge coupling constant.

Passing to the noncommutative Yang-Mills theory with $U_\star(N)$
gauge symmetry, we introduce the $u_\star(N)$ algebra, whose
elements can be expanded as
\be\label{un-algebra} f=\sum_{a=0}^{N^2-1}\ f^a(x) T^a\,, \ee
and the $u_\star(N)$ Lie-algebra is defined with the star-matrix
bracket:
\be\label{un-bracket} [f, h]_{\star}= f\star h - h \star f\;
,\;\;\;\;\;\  f,\,h\in u_{\star}(N)\,. \ee
The $U_\star(N)$ gauge theory is described by the
$u_\star(N)$-valued gauge fields
\be\label{un-g.f.} \hat A_{\mu}(x)=\sum_{a=0}^{N^2-1}\ \hat
A_{\mu}^a(x) T^a\,, \ee
with the noncommutative gauge transformations
\be\label{g.t.} \hat A_{\mu}\to\ \hat A'_{\mu}= \hat
A_{\mu}+\hat\delta_{\hat\Lambda} \hat A_{\mu}=\hat A_{\mu} +
\partial_{\mu} \hat\Lambda + ig [\hat\Lambda, \hat A_\mu]_{\star}\ ,
\ \; \hat\Lambda=\sum_{a=0}^{N^2-1}\ \hat\Lambda^a(x)\,T^a\,\in
u_{\star}(N)\,. \ee
The field strength is correspondingly defined with \sp,
\be\label{f-s} \hat F_{\mu\nu}=\partial_{[\mu}\hat A_{\nu]} + i
g[\hat A_{\mu},\hat A_{\nu}]_{\star}\,, \ee
and it transforms covariantly under the infinitesimal $u_\star(N)$
gauge transformations:
\be \hat F_{\mu\nu}\to\ \hat F'_{\mu\nu}= \hat F_{\mu\nu} +i g
[\hat\Lambda,\hat F_{\mu\nu}]_{\star}\,. \ee

One of the most striking peculiarities that the \sp\ introduces in
the structure of the gauge groups appears for the rank 1, where the
group \Uo\ shows a ``non-Abelian'' behaviour, unlike the ordinary
$U(1)$. This feature leads to the \textit{charge quantization
property} \cite{Hayakawa} in NC QED and in any other model involving
\Uo\ gauge fields coupled to matter, such as the noncommutative
versions of the \sm\ \cite{NCSM,Wess-NCSM}.

\subsection{Charge quantization
problem}\label{charge-quant-section}

The charge quantization problem arises from the non-Abelian
character of the gauge group \Uo\ and  the restriction on the
representations of \Uo to the (anti-)fundamental, adjoint and
singlet ones, corresponding respectively to $\pm 1$ and zero
charges. (For the adjoint representation, although the charge
vanishes, the fields have (electric) dipole moment, see e.g.
\cite{NCSM,Ihab}.) The Seiberg-Witten map connects the non-Abelian
\Uo\ gauge symmetry to the Abelian $U(1)$ symmetry, therefore the
question arises whether the charge quantization problem stands also
in this context.

It is well known that for a non-Abelian gauge group the charge is
fixed by the representation and two fields in the the same
representation can not couple to the same gauge boson with two
different coupling constants. Due to  its non-Abelian nature, the
\Uo\ gauge field can couple only with a given charge, and therefore
in a theory with more than one value for the (electric)
charge\footnote{The latter can be thought as \Uo\ theories with
different couplings, the ratio of which is equal to the ratio of the
charges in the problem.} we have to add in the model as many \Uo\
gauge fields as the number of  matter fields with different charges.
For instance, in the \sm we have to introduce as many noncommutative
hyperphotons as the number of hypercharges, which is six for the
Standard Model. This leads to an increase in the number of degrees
of freedom in the gauge sector, which has to be reduced
consistently, by \emph{spontaneous symmetry breaking}, to one single
hyperphoton. The question arises whether, after performing this
spontaneous symmetry breaking, the various matter fields will couple
with the appropriate hypercharges to the residual $U(1)$
hyperphoton.

We can take as a showcase the situation with only two different
noncommutative matter fields in (fundamental representation of)
$U_\star(1)$, with two different hypercharges: the field $\hat\Psi$,
with the hypercharge $q_1$, couples to the hyperphoton $\hat
A^1_\mu$, while the field $\hat\Phi$, with the hypercharge $q_2$,
couples to the hyperphoton $\hat A^2_\mu$:
\bea D_\mu\hat \Psi&=&\partial_\mu\hat\Psi-iq_1 \hat
A^1_\mu\star\hat\Psi\,,\cr
D_\mu\hat \Phi&=&\partial_\mu\hat\Phi-iq_2 \hat
A^2_\mu\star\hat\Phi\,. \eea
Using the \SW map (see Appendix \ref{SW-review-appendix}) the \nc
fields $\hat A_\mu^1$, $\hat A_\mu^2$, as well as $\hat\Psi$,
$\hat\Phi$, are expanded correspondingly in terms of ordinary $U(1)$
gauge fields, $A_\mu^1$ and $A_\nu^2$, and the ordinary matter
fields $\Psi$ and $\Phi$, namely,
\bea\label{A1A2-SW-map}%
\hat
A^1_\mu(A)&=&A^1_\mu-\frac{q_1}{4}\theta^{\sigma\tau}\{A^1_\sigma,\partial_\tau
A^1_\mu +F^1_{\tau\mu}\}+\ord{\theta^2}\,,\cr \hat
A^2_\mu(A)&=&A^2_\mu-\frac{q_2}{4}\theta^{\sigma\tau}\{A^2_\sigma,\partial_\tau
A^2_\mu +F^2_{\tau\mu}\}+\ord{\theta^2} \eea
and
\bea\label{Psi-Phi-SW-map}%
\hat\Psi(\Psi,A_\mu)=\Psi-\frac{q_1}{2}\theta^{\mu\nu}A^1_\mu\partial_\nu\Psi+{\cal
O}(\theta^2)\,,\cr
\hat\Phi(\Psi,A_\mu)=\Phi-\frac{q_2}{2}\theta^{\mu\nu}A^2_\mu\partial_\nu\Phi+{\cal
O}(\theta^2)\,, \eea
The covariant derivatives of the ordinary fields are:
\bea\label{Psi-Phi-cov-der}%
D_\mu \Psi&=&\partial_\mu\Psi-iq_1A^1_\mu\Psi\,,\cr D_\mu
\Phi&=&\partial_\mu\Phi-iq_2 A^2_\mu\Phi\,.\eea

 The ordinary gauge fields $A_\mu^1$ and $A_\mu^2$ can not be imposed to
coincide by hand. It should be emphasized that the properties of the
noncommutative gauge fields to which one maps the ordinary gauge
fields fix the form of the interaction in the ordinary action, i.e.
the types of ordinary gauge-invariant terms, as well as the
couplings. Since the noncommutative gauge field $\hat A_\mu^1$
couples to a matter field with the charge $q_1$, it means that in
the ordinary action the field $A_\mu^1$ will couple only with the
charge $q_1$, although, as Abelian gauge field, it has the
possibility to couple with any other charge. If we wish to couple
the ordinary gauge field with another coupling constant than the one
specified by the Seiberg-Witten map, the respective coupling will
obviously have no correspondent in the noncommutative action, and as
such it will have to be discarded.

We can now perform a spontaneous symmetry breaking of the ordinary
$U(1)\times U(1)$ symmetry, introducing a complex scalar field
$\chi$, with an appropriate Higgs potential leading to a vacuum
expectation value $\langle\chi\rangle=v$. As already mentioned, this
ordinary field can couple to $A_\mu^1$ only by $\pm q_1$ and to
$A_\mu^2$ only by $\pm q_2$. Moreover, it has to satisfy another
constraint of the no-go theorem, namely that it can couple
simultaneously to the two gauge fields only in the fundamental
representation of one and antifundamental representation of the
other, i.e.
\be D_\mu\chi=(\partial_\mu-iq_1 A^1_\mu+iq_2A^2_\mu)\,\chi.\ee
Using the Higgs mechanism, we obtain a mass term for the combination
\be\label{massive} {\cal
A}^1_\mu=\left(\frac{q_1}{\sqrt{q_1^2+q_2^2}}A^1_\mu-\frac{q_2}{\sqrt{q_1^2+q_2^2}}A^2_\mu\right),\
\ \ m^2_{{\cal A}}=\frac{v^2}{2}(q_1^2+q_2^2)\,,\ee
while the orthogonal combination
\be\label{massless} {\cal
A}^2_\mu=\left(\frac{q_2}{\sqrt{q_1^2+q_2^2}}A^1_\mu+\frac{q_1}{\sqrt{q_1^2+q_2^2}}A^2_\mu\right)\ee
remains massless -- the gauge field of the residual $U(1)$ symmetry,
i.e. the ``true" hyperphoton.

Inverting \eqref{massive} and \eqref{massless}, and introducing the
result into \eqref{Psi-Phi-cov-der}, we find the coupling of the
matter fields $\Psi$ and $\Phi$ to the residual gauge boson ${\cal
A}_\mu^2$:
\bea D_\mu \Psi&=&\partial_\mu\Psi-i\frac{q_1q_2}{\sqrt{q_1^2+q_2^2}}{\cal A}^2_\mu\Psi+\ldots\,,\\
D_\mu \Phi&=&\partial_\mu\Phi-i\frac{q_1q_2}{\sqrt{q_1^2+q_2^2}}
 {\cal A}^2_\mu\Phi+\ldots\,. \eea%
Thus, the two matter fields originally coupled with different
charges to the $U(1)$ gauge bosons, will eventually couple with the
same charge to the residual gauge boson ${\cal A}_\mu^2$, upon the
reduction of the degrees of freedom by spontaneous symmetry
breaking. It is straightforward to see that this result carries over
to the noncommutative action via the Seiberg-Witten maps
\eqref{A1A2-SW-map} and \eqref{Psi-Phi-SW-map}.

It should be mentioned that the desired couplings with charges $q_1$
and $q_2$, respectively, of the matter fields $\Psi$ and $\Phi$ to
the residual gauge bosons could have been obtained only if the
combination $A^1_\mu-A^2_\mu$ could be made massive. Arguing
backwards, this could have happened only if the Higgs field $\chi$
could couple with the same charge to both $A^1_\mu$ and $A^2_\mu$.
However, as emphasized above, such a coupling could not have been
taken by \SW map to the noncommutative action, consequently it is
not allowed either in the ordinary action.

The result is not surprising, since the residual $U(1)$ symmetry in
the language of ordinary fields has to be expressible as a
$U_\star(1)$ symmetry in the language of noncommutative fields. A
$U_\star(1)$ symmetry has the charge quantization problem, even if
it is a residual one, and the mechanism of spontaneous symmetry
breaking takes care of making equal the couplings of the matter
fields to the residual gauge boson.

\subsection{Maximal number of charges for a matter
field}\label{not-more-than-two-section}

For a gauge group of the form $\prod_{i=1}^n U_\star (N_i)$, as
stated by the no-go theorem, a matter field can be charged under at
most two of the group factors, $U_\star (N_i)$ and $U_\star (N_j)$,
the matter field being necessarily in the fundamental representation
of one while in the antifundamental of the other.\footnote{The only
known way of circumventing the no-go theorem in this respect is by
the use of noncommutative half-infinite Wilson lines attached to the
matter fields \cite{Gross}, in which case the matter fields can be
in tensor representations of any number of $U_\star(N_i)$ factors
\cite{Chu:2001if,Chu:2002,ASTU}.} The question we would like to
address in this section, namely checking if this restriction still
persists under the \SW map, in effect reduces to studying the
possibility of having a matter field in the fundamental
representations of two noncommutative gauge groups.

Let us assume that the field $\hat\Psi$ is in the fundamental
representation of $U_\star (N_1)$ and $U_\star(N_2)$:
\bea \hat\delta_{\hat\Lambda}
\hat\Psi&=&i\,g_1\hat\Lambda(\Lambda,A_\mu)\star\hat\Psi\,,\ \ \ \ \ \ A_\mu,\Lambda\in u(N_1),\ \ \hat\Lambda(\Lambda,A_\mu)\in u_\star(N_1)\,,\\
\hat\delta_{\hat\Sigma}
\hat\Psi&=&i\,g_2\hat\Sigma(\Sigma,B_\mu)\star\hat\Psi\,,\ \ \ \ \ \
B_\mu,\Sigma\in u(N_2),\ \ \hat\Sigma(\Sigma,B_\mu)\in
u_\star(N_2)\,, \eea
where $A_\mu$ and $B_\mu$ are the ordinary gauge fields and
$\Lambda$ and $\Sigma$ are the gauge parameters. The coupling
constants corresponding to the two gauge symmetries are $g_1$ and
$g_2$, respectively. Next, consider two successive transformations
of $\hat\Psi$, corresponding to the two symmetries and their
reverse, i.e.
$(\hat\delta_{\hat\Lambda}\hat\delta_{\hat\Sigma}-\hat\delta_{\hat\Sigma}\hat\delta_{\hat\Lambda})\hat\Psi(x)$.
In the spirit of the \SW map, we have to require the compatibility
between the ordinary and the noncommutative gauge transformations
\eqref{consistency1}. As two ordinary gauge transformations under
two different gauge algebras are independent, we shall require the
independence of the corresponding noncommutative gauge
transformations, i.e.
\be\label{indep}(\hat\delta_{\hat\Lambda}\hat\delta_{\hat\Sigma}-\hat\delta_{\hat\Sigma}\hat\delta_{\hat\Lambda})\hat\Psi(x)=0\,.\ee
The requirement \eqref{indep} is actually a particular case of
\eqref{consistency1}, since
\be [\Lambda,\Sigma]=0, \ \ \ \ \mbox{when} \  \ \ \Lambda\in
u(N_1),\ \ \Sigma\in u(N_2)\,.\ee
Thus, the full set of Seiberg-Witten map consistency conditions for
a noncommutative matter field in the fundamental representation of
two noncommutative $u_\star (N)$ algebras is:
\bea
(\hat\delta_{\hat\Lambda}\hat\delta_{\hat\Lambda'}-\hat\delta_{\hat\Lambda'}\hat\delta_{\hat\Lambda})\hat\Psi(x)
&=&\hat\delta_{ig_1\widehat{[\Lambda,\Lambda']}}\hat\Psi(x)\,,\ \ \
\ \ \
\Lambda,\Lambda'\in u(N_1),\label{g1}\\
(\hat\delta_{\hat\Sigma}\hat\delta_{\hat\Sigma'}-\hat\delta_{\hat\Sigma'}\hat\delta_{\hat\Sigma})\hat\Psi(x)
&=&\hat\delta_{ig_2\widehat{[\Sigma, \Sigma']}}\hat\Psi(x)\,,\ \ \ \
\
\ \Sigma, \Sigma'\in u(N_2),\label{g2}\\
(\hat\delta_{\hat\Lambda}\hat\delta_{\hat\Sigma}-\hat\delta_{\hat\Sigma}\hat\delta_{\hat\Lambda})\hat\Psi(x)&=&
0\,.\label{g1g2} \eea
The conditions \eqref{g1} and \eqref{g2} lead to the expressions for
$\hat\Lambda(\Lambda,A_\mu)$ and $\hat\Sigma(\Sigma,B_\mu)$:
\bea\label{Lambda-Sigma}
\hat\Lambda(\Lambda,A_\mu)=\Lambda+\frac{g_1}{4}\theta^{\mu\nu}\{\partial_\mu\Lambda,A_\nu\}+\ldots\,,\cr
\hat\Sigma(\Sigma,B_\mu)=\Sigma+\frac{g_2}{4}\theta^{\mu\nu}\{\partial_\mu\Sigma,B_\nu\}+\ldots\,,\eea
which have to be inserted into \eqref{g1g2}, rewritten as
\be\hat\Lambda(\Lambda,A_\mu)\star\hat\Sigma(\Sigma,B_\mu)-\hat\Sigma(\Sigma,B_\mu)\star\hat\Lambda(\Lambda,A_\mu)=0\,.\label{compat3}\ee
Up to the first order in $\theta$, the condition \eqref{compat3}
becomes
\bea\label{compat3'}
[\Lambda\,,\Sigma]\,+\,\frac{i}{2}\,\theta^{\mu\nu}
[\partial_\mu\Lambda\,,\partial_\nu\Sigma]+\,\frac{g_2}{4}\theta^{\mu\nu}\,
[\Lambda,\{\partial_\mu\Sigma,B_\nu\}]\
+\,\frac{g_1}{4}\theta^{\mu\nu}\,
[\{\partial_\mu\Lambda,A_\nu\},\Sigma]+\ord{\theta^2}=0\,.\eea
The last line in \eqref{compat3'} contains the commutators $[a,b]$
with $a\in u(N_1)$ and $b\in u(N_2)$. Besides the Moyal bracket of
$\Lambda$ and $\Sigma$, only such ordinary commutators will appear
in all orders in $\th$, and all of them will vanish. Consequently,
\bea\label{inconsist}
(\hat{\delta}_{\hat{\Lambda}}\,\hat{\delta}_{\hat{\Sigma}}\ -\
\hat{\delta}_{\hat{\Sigma}}\,\hat{\delta}_{\hat{\Lambda}})\hat
\Psi(x)\
&=&-g_1g_2\left([\Lambda\,,\Sigma]\,+\frac{i}{2}\theta^{\mu\nu}[\partial_\mu
\Lambda,
\partial_\nu\Sigma]+\ord{\theta^2}\right)\star\hat\Psi(x)\cr
&=&-g_1g_2\,[\Lambda\,,\Sigma]_\star\star\hat\Psi(x)\,,\eea
which is valid in all orders in $\theta$. The r.h.s. of the
expression \eqref{inconsist} can not vanish, due to the \sp\ in the
Moyal bracket of the gauge parameters belonging to the two
independent gauge algebras. This shows that the equations
\eqref{g1}-\eqref{g1g2} can not be simultaneously fulfilled,
consequently a noncommutative matter field can not be in the
fundamental representation of more than one gauge symmetry. We point
out that, for a
   field transforming in
   bi-fundamental representation of  $U_\star(N_1) \times U_\star(N_2)$
   (i.e. fundamental of  $U_\star(N_1)$ and antifundamental of $U_\star(N_2)$), the condition \eqref{g1g2} is naturally satisfied.

\subsection{String theory considerations}\label{string-theory}

It is instructive to support the above algebraic arguments by string
theory reasoning. Indeed, in the string theory setting it is clear
that the \SW map can only be worked out when the no-go theorem
holds. To see this, it is enough to revisit the no-go theorem, the
\SW map and their connection from the string theory viewpoint. The
NC Moyal plane can be obtained as the low energy effective theory of
D-branes in the presence of a Neveu-Schwartz two-form $B$-field
\cite{NC-Branes}. This is a particular $\alpha'\to 0$ limit, while a
certain combination of the background $B$-field and $\alpha'$, which
appears as the noncommutativity parameter $\th$, is held fixed
\cite{SW}. The Moyal plane is what is seen from such a D-brane when
probed by \emph{open strings} whose end-points are attached to the
brane. The first part of the no-go theorem in this picture is
related to the simple fact that an open string has two end-points.
To see this, we note that when we have $N$ D-branes on top of each
other, one should then associate a $U(N)$ Chan-Paton factor to each
end, i.e. if one end is in the fundamental representation of $U(N)$,
the other end should necessarily be in the antifundamental
representation \cite{Witten}. (Recall that open strings ending on
D-branes are orientable.) This property persists when we have a
constant background $B$-field, i.e. the end-points of open strings
are in fundamental and antifundamental representations of
$U_\star(N)$ and the whole string is then in the adjoint
representation of $U_\star(N)$.

To argue for the second part of the no-go theorem, namely the
not-more-than-two charges, we need to consider a D-brane setup which
realizes the $\prod_{i=1}^nU_\star(N_i)$ gauge group. This is done
if we consider $n$ stacks of parallel D-branes, each consisting of
$N_i,\ i=1,\ldots, n$ branes, while the stacks are separated by
distances much larger than the string scale. One can then recognize
two classes of open strings: those which have both ends on the same
stack and those which are stretched between different stacks. The
former give rise to the $U_\star(N_i)$ gauge fields, while the
latter lead to matter fields in ``bi-fundamental'' representation of
$U_\star (N_i)\times U_\star (N_j),\ i\neq j$. Again, since the open
string has two end-points and open strings stretched between
D-branes are orientable, there is no room for Chan-Paton factors for
the product of more than two $U_\star(N_i)$.

   The \SW map, on the other hand, as explained in \cite{SW},
comes from the equivalence of two pictures for the dynamics of
$D$-branes in the background $B$-field. Explicitly, one can have two
different descriptions, the open string description and the closed
string description. The open string description is the
noncommutative one and the closed string description is the ordinary
commutative one. The \SW map in this language expresses the
equivalence of the Born-Infeld actions in these two descriptions. If
we take this as an equivalent definition for the \SW map, then it
clearly states that the \SW map only works for the cases where the
noncommutative description in terms of open strings exists, namely
the cases which also respect the no-go theorem.

\section{Concluding remarks and
discussion}\label{discussion-section}

We have shown that the \SW map can only be used for the cases which
respect the noncommutative gauge theory no-go theorem. In
particular, we have been dealing with $U_\star(N)$ gauge groups or a
product of them. On the other hand, using the \SW map as the
defining relation, it has been argued that one can construct a
noncommutative version of Lie groups/algebras other than $U(N)$
\cite{Madore,Wess-enveloping,Wess-general-NCgauge-theory}. This, as
reviewed in the Appendix \ref{enveloping-appendix}, is achieved by
allowing the gauge field to take values in the enveloping algebra of
the gauge algebra ${\cal A},\ {\cal U}({\cal A})_\star$, with a
combination fixed by the \SW map. The particular case of obvious
interest is when ${\cal A}=su(N)$.

One may then wonder whether the arguments of Subsection
\ref{not-more-than-two-section} would still hold for products of
${\cal U}({\cal A}_i)_\star$. Recalling the discussions reviewed in
the Appendices, the key equations to be considered are again
\eqref{g1}-\eqref{g1g2}, but now $\Lambda, \Lambda'$ and $\Sigma,
\Sigma'$ take values in the enveloping algebras ${\cal U}({\cal
A}_1)_\star$ and ${\cal U}({\cal A}_2)_\star$, respectively. Eq.
\eqref{Lambda-Sigma} is still a solution to \eqref{g1} and
\eqref{g2}, while \eqref{g1g2} reduces to \eqref{compat3}. Again we
face the same problem as in the $u_\star(N)$ case and therefore we
conclude that one can not have matter fields which are in the
fundamental representation of more than one ${\cal U}({\cal
A})_\star$ enveloping algebra.

Having shown the persistence of the NC gauge theory no-go theorem
within the \SW map expansion and its extension to the cases with the
NC gauge theories based on the enveloping algebras, we would like to
comment on the physical implications of this result for particle
physics model building.

There have been two different proposals for building a NC version of
the SM. One which has been proposed by the authors in \cite{NCSM} is
based on the $U_\star(3)\times U_\star(2)\times U_\star(1)$ gauge
theory. The matter content is chosen in such a way that it respects
the NC gauge theory no-go theorem. Indeed, with this gauge group the
most general possible matter content is exactly the one which has
been realized in Nature, within the SM. In this sense we have the
remarkable property that the matter content is completely specified:
fixing the NC gauge group fixes the matter content. With this gauge
group, the theory has two extra gauge fields compared to the SM. To
reduce this theory to the SM, at least at scales below TeV, we
employed the  ``Higgsac symmetry reduction mechanism'' to give
masses to these two extra gauge fields. This symmetry reduction
mechanism is not spontaneous, but a solution to this problem is
under consideration \cite{workinprogress}. Besides this point, this
model, as it stands and was presented in \cite{NCSM}, suffers from
the chiral anomaly \cite{Carmelo} (a suggestion to cure this problem
was made in \cite{CKT}).

The other proposed formulation for the NC version of the SM is the
one based on  ${\cal U} (su(3))_\star\times{\cal U}
(su(2))_\star\times {\cal U}(u(1))_\star$ \cite{Wess-NCSM}, where
the applicability of the \SWmap has been assumed and used. Although
this model has the same particle and gauge field content as the SM,
it suffers from other problems: in order for that proposal to work,
the NC gauge theory no-go theorem should be bypassed. This is easy
to see if we recall that \emph{i)} in the SM there are six different
hypercharges, and this contradicts the \Uo charge quantization and
\emph{ii)} the left-handed doublet of quarks is charged under all
three gauge symmetries. However, as we have argued here, the no-go
theorem
 still applies despite the usage of the \SW
map, upon which that proposal is built.

Therefore, it appears that both of the NC SM proposed have model
building problems and constructing a NC version of the SM still
remains a challenge. Finding other representations which have not
been accounted for in the no-go theorem, in particular the existence
of the $N$-dimensional totally antisymmetric representation of
${\cal U}(u(N))_\star$, i.e. the representations under the
$u_\star(1)$ part of $u_\star(N)$, may prove useful in this respect.

This challenge may be an indication for another possibility that the
NC effects could only become important at energies much above the
TeV scale and that one should think of a NC GUT rather than the NC
SM.


\section*{Acknowledgements}

M.M. Sh.-J. would like to thank the High Energy, Cosmology and
Astroparticle Physics Section of the Abdus Salam ICTP and the
Elementary Particle Physics Division of the Department of Physics,
University of Helsinki for their hospitality when this project
started. The support of the Academy of Finland under the Projects
No. 121720 and 127626 is greatly acknowledged.

\appendix
\section{Noncommutative gauge theories and Seiberg-Witten map}\label{SW-review-appendix}

The \SW map \cite{SW}, as originally proposed, is a map between the
NC $U_\star(N)$ gauge fields and gauge transformations, respectively
denoted by $\hat A$ and $\hat \Lambda$, and the corresponding
ordinary $u(N)$-matrix valued functions $A$ and $\Lambda$.

The \SW map is a field redefinition of the form%
\be\label{SW-gauge-field}%
\hat A = \hat A(A, \partial A,\partial^2 A,\ldots;\theta), %
\ee%
accompanied by a reparametrization
\be\label{SW-gauge-trans}%
\hat\Lambda=\hat\Lambda(\Lambda, \partial \Lambda, \partial A,
\partial^2\Lambda, \partial^2 A, \ldots;\theta).\ee
The map of the ordinary gauge field $A$ to the noncommutative gauge
field $\hat A$ is required to preserve the gauge equivalence
relation, though the groups of the noncommutative and the ordinary
gauge theories are different. This can be acquired by allowing $\hat
\Lambda$ to depend on both $\Lambda$ and $A$. The requirement of
identification of the gauge equivalence is concisely written as:
\be\label{SWmap}\hat A(A)+\hat\delta_{\hat\Lambda}\hat A(A)=\hat
A(A+\delta_\Lambda A),\ee
with infinitesimal $\Lambda$ and $\hat\Lambda$. The solutions of
\eqref{SWmap} for $\hat A$ and $\hat \Lambda$ are assumed to be
\emph{smooth and local} to all orders in $\theta$. Eq. \eqref{SWmap}
can be solved as a series expansion in powers of $\theta$. To first
order, the map reads \cite{SW}:
\bea \label{SW-map-solution-gauge-field}\hat
A_\mu(A)&=&A_\mu-\frac{g}{4}\theta^{\sigma\tau}\{A_\sigma,\partial_\tau
A_\mu +F_{\tau\mu}\}+\ord{\theta^2}\,,\cr %
\hat\Lambda(\Lambda,A)&=&\Lambda+\frac{g}{4}\theta^{\mu\nu}\{\partial_\mu\Lambda,A_\nu\}+\ord{\theta^2}\
,\eea%
where  $\{\cdot, \cdot\}$ is the anticommutator of $u(N)$-valued
matrices.

The \SW map can be extended beyond the pure gauge theory. For the NC
matter field $\hat \Psi$ in the fundamental (or antifundamental)
representation of $u_\star(N)$,
with the gauge transformation rule%
\be\label{matter-gauge-trans}%
\hat\delta_{\hat\Lambda} \hat\Psi=ig\,\hat\Lambda\star \hat\Psi\ . %
\ee%
and the corresponding ordinary counterpart
$\delta\Psi=ig\Lambda\Psi$, the \SW map is defined by%
\be\label{SW-map-matter-def}%
\hat\delta_{\hat\Lambda}\hat\Psi=\delta_\Lambda\hat\Psi,%
\ee%
where in the r.h.s. of the above equation $\hat \Psi$ should be considered as
\be\label{matter-SW-map}%
\hat\Psi=\hat\Psi (\Psi, A; \theta) %
\ee%
and $\delta_\Lambda$ means the ordinary gauge transformation of the
ordinary fields on which $\hat\Psi$ depends. (Note that
\eqref{matter-gauge-trans} implies that \eqref{matter-SW-map} should
be linear in $\Psi$ and its derivatives.) The above equation, too,
can be solved together with \eqref{SWmap} as a systematic power
series expansion in $\theta$. In the first order the map reads
\be\label{SW-map-solution-matter}%
\hat\Psi(\Psi,A)=\Psi-\frac{g}{2}\theta^{\mu\nu}A_\mu\partial_\nu\Psi+{\cal
O}(\theta^2)\ . \ee%

In order that \eqref{SW-map-matter-def} has a solution, one should
require the closure of the \SW map under the commutator of
successive gauge transformations,
\be\label{consistency1}%
(\hat\delta_{\hat\Lambda}\hat\delta_{\hat\Sigma}-\hat\delta_{\hat\Sigma}\hat\delta_{\hat\Lambda})\hat\Psi(x)
=\hat\delta_{ig\,\widehat{[\Lambda,\Sigma]}}\hat\Psi(x)\,,%
\ee%
which leads to the following \emph{consistency} condition%
\be\label{consistency2}%
i\delta_\Lambda\hat\Sigma(\Sigma,A)-i\delta_\Sigma\hat\Lambda(\Lambda,A)
-g\,\hat\Lambda(\Lambda,A)\star\hat\Sigma(\Sigma,A)+g\,\hat\Sigma(\Sigma,A)\star\hat\Lambda(\Lambda,A)=i\hat\Omega(\Omega,
A)\,,\ee%
where $\hat\Omega(\Omega, A)$ is the gauge parameter of the NC gauge
transformation for $\Omega=ig\,[\Lambda,\Sigma]$ and the variations
of the type $\delta_\Lambda\hat\Sigma(\Sigma,A)$ refer to the
ordinary variation $\delta_\Lambda$ of the gauge field $A$ on which
$\hat\Sigma$ depends.

\section{Enveloping algebra valued noncommutative gauge
fields and the Seiberg-Witten map}\label{enveloping-appendix}

The \SW map can be used to construct gauge field theories other than
$U_\star(N)$ theory \cite{Wess-enveloping,
Wess-general-NCgauge-theory}. Here we sketch this construction. To
start with, consider the definition of the \SW map, \eqref{SWmap}
and \eqref{SW-map-matter-def}, and let the ordinary gauge fields
take values in the Lie algebra ${\cal A}$. Assume that these fields
are in some \nbyn\ matrix form of ${\cal A}$. At zeroth order in
$\theta$,  $\hat A_\mu$, $\hat\Lambda$ and $\hat\Psi$ are equal to
their ordinary counterparts, hence transforming as representations
of the algebra ${\cal A}$. It is readily seen that the solutions to
\eqref{SWmap} and \eqref{SW-map-matter-def} at first order in
$\theta$ for a generic algebra ${\cal A}$ are again of the form
\eqref{SW-map-solution-gauge-field} and
\eqref{SW-map-solution-matter}. These expressions are not taking
values in the algebra ${\cal A}$ (except for ${\cal A}=u(N)$), but
they are in general in the (universal) enveloping algebra of ${\cal
A}$, i.e. ${\cal U}({\cal A})$, with the specific expression which
is fixed by the \SW map and will be denoted by ${\cal U}({\cal
A})_\star$. If we allow the fields to fall into the representations
of ${\cal U}({\cal A})_\star$ (rather than ${\cal A}$)\footnote{Note
that the ``gauge invariant'' physical observables are also
constructed from quantities in ${\cal U}({\cal A})_\star$ and they
involve the parts of the \SW map which are not ${\cal A}$-valued.}
this will show a way for constructing NC gauge theories based on the
algebra ${\cal A}$. In this construction the zeroth order NC fields
are taken to be in ${\cal A}$ and although the defining equations
\eqref{SWmap} and \eqref{SW-map-matter-def} have a closed form, the
explicit form of the action for these theories can only be given as
a power series expansion in $\theta$.

The above can be made more explicit if we choose a specific basis
for ${\cal A}$. To illustrate the idea, let us focus on the
phenomenologically interesting case of ${\cal A}=su(N)$ for which,
in the notations of the Subsection \ref{ord_vs_NC}, the covariant
derivative of the field $\Psi$ is
\be D_\mu\Psi(x)=(\partial_\mu-igA_\mu(x))\Psi(x),\ee
with the gauge field $A_\mu(x)$ transforming as in \eqref{g.t.ord},
but this time with $\Lambda=\sum_{a}\,\Lambda^a(x)\,T^a\,\in su(N)$,
$a=1,2,\ldots ,N^2-1$. The noncommutative gauge theory to which one
maps the above ordinary gauge theory will be invariant under the
$\star$-transformations valued in the enveloping algebra ${\cal U}
(su(N))_\star$, i.e. a matter field will transform as:
\be\label{env-gt}\hat\delta_{\hat\Lambda}\hat\Psi(x)=ig\,\hat\Lambda(x)\star\hat\Psi(x),\ee
with
\begin{equation}
\label{9} \hat\Lambda(x) = \hat\Lambda_a(x)T^a +
\hat\Lambda^1_{ab}(x) : T^a T^b: + \ldots + \hat\Lambda^{n-1}_{a_1
  \dots a_{n}}(x) : T^{a_{1}}\dots T^{a_n} : + \ldots\,,
\end{equation}
where the dots indicate summation over a basis of ${\cal U}
(su(N))_\star$, for example the basis of completely symmetrized
products of the Lie algebra:
\begin{eqnarray}
\label{env-basis}
:T^a: & =& T^a\,, \\
:T^aT^b: & = & \frac{1}{2}\{T^a,T^b\}= \frac{1}{2}(T^aT^b+T^bT^a)\,,\nonumber \\
:T^{a_1}\dots T^{a_n}: & = & \frac{1}{n!}\sum_{\pi \in
S_n}T^{a_{\pi(1)}}\dots T^{a_{\pi(n)}},\nonumber
\end{eqnarray}%
and
\begin{equation}
\label{16a} \hat\Lambda(\Lambda,A) = \Lambda
+\frac{g}{4}\theta^{\mu\nu}\{\partial_\mu\Lambda,A_\nu\}+{\cal
O}(\theta^2)
=\Lambda+\frac{g}{2}\theta^{\mu\nu}\partial_\mu\Lambda^a A_{\nu}^b
:T^aT^b:+{\cal O}(\theta^2) ,
\end{equation}%
and similarly for $\hat A$ and $\hat\Psi$. The higher orders of the
expansions are obtained analogously. In
\cite{Wess-general-NCgauge-theory} the action of a noncommutative
gauge theory with fermionic matter has been derived to the second
order in the noncommutativity parameter $\theta$, and the result
being written only in terms of ordinary gauge covariant derivatives
and field strengths, exhibits beautifully the ordinary gauge
invariance of the expansion.

However, one may wonder about the consistent use of the above
definition for the ${\cal U}({\cal A})_\star$ theory, in the case of
model building, when several ${\cal U}({\cal A}_i)_\star$ factors
are present. As in the $u_\star(N)$ case, one can work out the
``integrability condition'' for the equations \eqref{SWmap} and
\eqref{SW-map-matter-def} by considering two successive gauge
transformations. This again leads to \eqref{consistency2} but now
the hatted objects take values in ${\cal U}({\cal A})_\star$. When
we have several gauge group factors, like $\prod_{i=1}^n {\cal
U}({\cal A}_i)_\star$, one should require the consistency condition
for each of the ${\cal U}({\cal A}_i)_\star$ factors. Moreover, as
discussed in Subsection \ref{not-more-than-two-section}, we need to
examine the consistency relations of the map for matter fields
charged under more than two gauge algebra factors -- and the result
is similar to the one obtained in Subsection
\ref{not-more-than-two-section}, i.e. matter fields transforming
under more than two noncommutative enveloping algebras can {\it not}
be constructed in this way.


\end{document}